\documentclass[aps,prb,twocolumn,showpacs,amsmath,amssymb]{revtex4}
\usepackage{graphicx}
\usepackage{bm}

\def\br{ \bm{r} }
\def\bk{ \bm{k} }
\def\bq{ \bm{q} }
\def\bR{ \bm{R} }
\def\bH{ \bm{H} }
\def\bK{ \bm{K} }
\def\bA{ \bm{A} }
\def\bD{ \bm{D} }
\def\bgam{ \bm{\gamma} }
\def\im{ \mathrm{Im} }
\def\re{ \mathrm{Re} }
\def\sign{ \,\mathrm{sign}\, }
\def\tr{ \,\mathrm{tr}\, }
\def\bnab{ \bm{\nabla}}
\def\bsigma{ \hat{\bm{\sigma}} }

\begin{document}

\title{Upper critical field in noncentrosymmetric superconductors}

\author{K. V. Samokhin}

\affiliation{Department of Physics, Brock University,
St.Catharines, Ontario L2S 3A1, Canada}
\date{\today}

\begin{abstract}
We calculate the upper critical field in superconductors without
inversion symmetry at arbitrary temperatures, in the presence of
scalar impurities. Both orbital and spin (paramagnetic) mechanisms
of pair breaking are considered. The superconducting phase
transition at nonzero field occurs into a helical vortex state, in
which the order parameter is modulated along the applied field (in
a cubic crystal). The helical state is stable with respect to
disorder. However, if the difference between the densities of
states in the electron bands split by spin-orbit coupling is
neglected, then the order parameter modulation is present only at
low temperatures, and only if the system is sufficiently clean and
paramagnetically limited. This state resembles the
Larkin-Ovchinnikov-Fulde-Ferrell state in centrosymmetric
superconductors.
\end{abstract}


\maketitle

\section{Introduction}
\label{sec: Intro}

Recently, superconductivity has been discovered in a number of
noncentrosymmetric compounds. Physical properties of these
materials vary considerably, from CePt$_3$Si (Ref.
\onlinecite{Bauer04}), in which strong electron correlations are
responsible for a heavy-fermion behavior, and the
superconductivity is likely anisotropic with gap nodes, to
Li$_2$Pd$_3$B and Li$_2$Pt$_3$B (Ref. \onlinecite{LiPt-PdB}),
which show rather conventional features, described by the
Bardeen-Cooper-Schrieffer (BCS) theory of phonon-mediated pairing.

The absence of inversion symmetry in the crystal lattice brings
about important qualitative changes, both in normal and
superconducting properties, compared with the centrosymmetric
case. These differences are highlighted, in particular, by
different responses to an external magnetic field. Distinctive
features of noncentrosymmetric superconductors include a strongly
anisotropic spin susceptibility with a large residual
component,\cite{Edel89,GR01,Yip02,FAKS04,FAS04,Sam05,Sam07}
magnetoelectric effect,\cite{Lev85,Edel89,Edel95,Fuji05} and novel
nonuniform (``helical'') superconducting
states.\cite{Agter03,Sam04,KAS05}

In this paper, we extend the classic theory of the upper critical
field, $H_{c2}(T)$, in BCS superconductors, which was developed in
Refs. \onlinecite{HW66} and \onlinecite{WHH66}, see also Ref.
\onlinecite{Scho91}, to the noncentrosymmetric case. The
spin-orbit (SO) coupling of electrons with a noncentrosymmetric
crystal lattice considerably changes the nature of single-electron
states, lifting spin degeneracy of the energy bands. This modifies
the Zeeman coupling of band electrons with magnetic field, which
in turn affects the paramagnetic pair breaking. The magnetic phase
diagram of noncentrosymmetric superconductors has been previously
studied in a two-dimensional case,\cite{BG02,DF07} in which the
orbital effects are not important. In Ref. \onlinecite{KAS05}, the
upper critical field for a three-dimensional Rashba superconductor
was calculated in the absence of impurities. The effects of
disorder on $H_{c2}$ in the Ginzburg-Landau regime near the
zero-field critical temperature have been investigated in Ref.
\onlinecite{MS07} for an arbitrary pairing symmetry, but without
paramagnetic effects. In this paper, we include both orbital and
paramagnetic mechanisms of pair breaking, as well as scalar
disorder, at arbitrary temperatures. We consider a
``minimal'' model of a three-dimensional weak-coupling
superconductor of cubic symmetry, with an isotropic
parabolic band split in two by the SO coupling. We assume that the 
ratio of the SO band splitting to the Fermi energy is small (which is the
case in most known noncentrosymmetric materials), and neglect the
impurity-induced mixing of the singlet and triplet components of
the gap function.

The paper is organized as follows: In Sec. \ref{sec: Model}, we
derive the equations for $H_{c2}(T)$, relegating some of the
technical details to Appendix. In Sec. \ref{sec: GL regime}, the
transition temperature as a function of the external field is
calculated in the Ginzburg-Landau region near the zero-field
critical temperature $T_{c0}$. In Sec. \ref{sec: Paramagnetic}, we
consider the purely paramagnetic limit, both in the clean and
disordered cases. The general case, with the orbital pair breaking
included, is studied in Sec. \ref{sec: General case}. Sec.
\ref{sec: Conclusions} contains a discussion of our results.
Throughout the paper we use the units in which $\hbar=k_B=1$.

\section{Basic equations}
\label{sec: Model}

Let us consider a noncentrosymmetric superconductor with the
Hamiltonian given by $H=H_0+H_{imp}+H_{int}$. The first term,
\begin{equation}
\label{H_band}
    H_0=\sum\limits_{\bk}[\epsilon_0(\bk)\delta_{\alpha\beta}+
    \bgam(\bk)\bm{\sigma}_{\alpha\beta}]
    a^\dagger_{\bk\alpha}a_{\bk\beta},
\end{equation}
describes non-interacting electrons in the crystal lattice
potential, where $\alpha,\beta=\uparrow,\downarrow$ are spin
indices, $\epsilon_0(\bk)$ is the quasiparticle energy, and
$\hat{\bm{\sigma}}$ are the Pauli matrices. We assume a parabolic
band and include the chemical potential in the dispersion
function: $\epsilon_0(\bk)=\bk^2/2m^*-\epsilon_F$, where $m^*$ is
the effective mass, $\epsilon_F=k_0^2/2m^*$, and $k_0$ is the
Fermi wave vector in the absence of the SO coupling. In Eq.
(\ref{H_band}) and everywhere below, summation over repeated spin
indices is implied, while summation over band indices is always
shown explicitly.

The second term in Eq. (\ref{H_band}) describes a Rashba-type SO
coupling of electrons with the crystal lattice.\cite{Rashba60} We
focus on the case of a noncentrosymmetric cubic crystal with the
point group $\mathbb{G}=\mathbf{O}$, which is applicable, for
instance, to Li$_2$(Pd$_{1-x}$,Pt$_x$)$_3$B. The simplest
expression for the SO coupling compatible with all symmetry
requirements has the following form:
\begin{equation}
\label{gamma_O}
    \bgam(\bk)=\gamma_0\bk,
\end{equation}
where $\gamma_0$ is a constant. It is convenient to characterize the
SO coupling strength by a dimensionless parameter
\begin{equation}
\label{SOC strength}
    \varrho=\frac{m^*|\gamma_0|}{k_0}.
\end{equation}
Diagonalization of the Hamiltonian (\ref{H_band}) yields two
non-degenerate bands labelled by helicity $\lambda=\pm$, which are
described by the following dispersion functions:
\begin{equation}
\label{xis}
    \xi_\lambda(\bk)=\epsilon_0(\bk)+\lambda|\bgam(\bk)|=
    \frac{k^2-k_0^2}{2m^*}+\lambda|\gamma_0|k.
\end{equation}
The SO band splitting is given by $E_{SO}=2|\gamma_0|k_0$, therefore 
$\varrho=E_{SO}/4\epsilon_F$.
While the two Fermi surfaces, defined by the equations
$\xi_\lambda(\bk)=0$, have different radii:
$k_{F,\lambda}=k_0(\sqrt{1+\varrho^2}-\lambda\varrho)$, the Fermi
velocities are the same: $\bm{v}_\lambda=v_F\hat{\bk}$, where
$v_F=(k_0/m^*)\sqrt{1+\varrho^2}$.

Scattering of electrons at isotropic scalar impurities is
introduced according to
\begin{equation}
\label{H_imp}
    H_{imp}=\int d^3\br\,
    U(\br)\psi^\dagger_\alpha(\br)\psi_\alpha(\br),
\end{equation}
where the impurity potential $U(\br)$ is a random function with
zero mean and the correlator $\langle
U(\br_1)U(\br_2)\rangle=n_{imp}U_0^2\delta(\br_1-\br_2)$,
$n_{imp}$ is the impurity concentration, and $U_0$ has the meaning
of the strength of an individual point-like impurity. The field
operators have the usual form: $\psi_\alpha(\br)={\cal
V}^{-1/2}\sum_{\bk}e^{i\bk\br}a_{\bk\alpha}$, where ${\cal V}$ is
the system volume.

We assume that there is a local attraction between electrons, e.g.
due to phonons, and describe the Cooper pairing by a BCS-like
Hamiltonian:
\begin{equation}
\label{H_int}
    H_{int}=-V\int d^3\br\,\psi_\uparrow^\dagger(\br)\psi_\downarrow^\dagger(\br)
    \psi_\downarrow(\br)\psi_\uparrow(\br),
\end{equation}
where $V>0$ is the coupling constant. A detailed analysis of the
relation between the microscopic pairing interaction and the gap
symmetry in the band representation can be found in Ref.
\onlinecite{SM08}. In particular, in the BCS-contact model
(\ref{H_int}) the local pairing of electrons with opposite spins
translates into the same-helicity pairing, the superconducting gap
function has only intraband components, and the order parameter is
represented by a single complex function $\eta(\br)$.

External magnetic field can be included in the electron band
theory by making the so-called Peierls substitution\cite{LL9} in
the dispersion functions $\xi_\lambda(\bk)$:
\begin{equation}
\label{k to K}
    \bk\to\bK=-i\frac{\partial}{\partial\br}+\frac{e}{c}\bA(\br),
\end{equation}
where $e$ is the absolute value of the electron charge and
$\bA(\br)$ is the magnetic vector potential. Near the upper
critical field, the magnetic induction is uniform,
$\bm{B}(\br)=\bH$, so that $\bA(\br)=(\bH\times\br)/2$ in the
symmetric gauge. This approach has been used to microscopically
derive the Ginzburg-Landau functional for noncentrosymmetric
superconductors, both in the clean\cite{Sam04} and
disordered\cite{MS07} cases. However, the bands (\ref{xis}) are
nonanalytic in $\bk$, and the Peierls substitution leads to
ill-defined operators. To avoid this, one can make the
substitution (\ref{k to K}) directly in the original Hamiltonian
[Eq. (\ref{H_band})]. Then, the SO coupling, the impurity
scattering, and the magnetic field are all incorporated in the
following single-electron Hamiltonian in the coordinate-spin
representation:
\begin{equation}
\label{hat h}
    \hat{h}=\frac{\bK^2}{2m^*}+\gamma_0\bK\bsigma
    +\frac{g}{2}\mu_B\bH\bsigma+U(\br)-\epsilon_F,
\end{equation}
where $g$ is the Land$\mathrm{\acute{e}}$ factor, and $\mu_B$ is
the Bohr magneton. The final expressions for observable
properties, in particular the upper critical field, do not
actually depend on whether the Peierls substitution is made in the
spin or band representations.

\subsection{Gap equations}
\label{sec: gap equation}

Following the standard textbook procedure, one can calculate the
difference between the disorder-averaged free energies in the
superconducting and normal states, at the same temperature and
field. In the vicinity of the upper critical field $H_{c2}(T)$,
one can keep only the terms quadratic in the order parameter in
the free-energy expansion:
\begin{equation}
\label{F gen}
    {\cal F}=\int\int
    d\br_1d\br_2\,\eta^*(\br_1)S(\br_1,\br_2)\eta(\br_2).
\end{equation}
We assume that the superconducting phase transition is of second
order. It should be noted that the validity of this assumption at
all temperatures is not obvious, see, e.g., a discussion in Sec.
\ref{sec: FOPT} below, and should be verified by examining the
higher-order terms in the free energy, which is beyond the scope
of the present work. The kernel in Eq. (\ref{F gen}) has the
following form:
\begin{eqnarray}
\label{S gen}
    S(\br_1,\br_2)=\frac{1}{V}\delta(\br_1-\br_2)-T\sum_n{}'
    X(\br_1,\br_2;\omega_n),
\end{eqnarray}
where
\begin{eqnarray}
\label{X def}
    &&X(\br_1,\br_2;\omega_n)=\frac{1}{2}g^\dagger_{\alpha\beta}g_{\gamma\delta}
    \nonumber\\
    &&\quad\times\bigl\langle G_{\beta\gamma}(\br_1,\br_2;\omega_n)
    G_{\alpha\delta}(\br_1,\br_2;-\omega_n)\bigr\rangle_{imp},
\end{eqnarray}
$\omega_n=(2n+1)\pi T$ is the fermionic Matsubara frequency, and
$\hat g=i\hat\sigma_2$. The prime in the second term in Eq.
(\ref{S gen}) means that the summation is limited to
$|\omega_n|\leq\omega_c$, where $\omega_c$ is the BCS frequency
cutoff. The angular brackets denote the impurity averaging, and
$\hat G(\br_1,\br_2;\omega_n)$ is the Matsubara Green's function
of electrons in the normal state, which satisfies the equation
\begin{equation}
\label{G eq gen}
    (i\omega_n-\hat{h}_1)\hat
    G(\br_1,\br_2;\omega_n)=\delta(\br_1-\br_2),
\end{equation}
with the Hamiltonian $\hat{h}$ given by expression (\ref{hat h}).
The subscript ``1'' means that $\hat{h}$ acts on the first
argument of the Green's function. The critical temperature at a
given field, or inversely the critical field at a given
temperature, is found from the condition that the lowest
eigenvalue of the operator $\hat S$, defined by the kernel (\ref{S
gen}), is zero.

At zero field, Eq. (\ref{G eq gen}) yields the following
expression for the average Green's function:\cite{MS07}
\begin{equation}
\label{G clean zero H}
    \hat{G}(\bk,\omega_n)=
    \sum_{\lambda=\pm}\hat\Pi_\lambda(\bk)G_\lambda(\bk,\omega_n),
\end{equation}
where
\begin{equation}
\label{Pis}
    \hat\Pi_\lambda(\bk)=\frac{1+\lambda\hat\bk\bsigma}{2}
\end{equation}
are the band projection operators ($\hat\bk=\bk/|\bk|$), and
\begin{equation}
\label{G lambda average zero H}
    G_\lambda(\bk,\omega_n)=
    \frac{1}{i\omega_n-\xi_\lambda(\bk)+i\Gamma\sign\omega_n},
\end{equation}
are the electron Green's functions in the band representation.
Here $\Gamma=1/2\tau$ is the elastic scattering rate, $\tau=(2\pi
n_{imp}U_0^2N_F)^{-1}$ is the electron mean free time due to
impurities, $N_F=(N_++N_-)/2$, and $N_\lambda$ is the Fermi-level
density of states in the $\lambda$th band.

Using the standard techniques,\cite{AGD} the impurity average of
the product of the two Green's functions in Eq. (\ref{X def}) can
be represented graphically by ladder diagrams, see Fig. \ref{fig:
ladder diagrams} (we assume the disorder to be sufficiently weak
for the diagrams with crossed impurity lines to be negligible).
Summation of the diagrams is facilitated by representing the
impurity line (which corresponds to each ``rung'' of the ladder)
as a sum of spin-singlet and spin-triplet terms:
\begin{equation}
\label{impurity line}
    n_{imp}U_0^2\delta_{\mu\nu}\delta_{\rho\sigma}=
    \frac{1}{2}n_{imp}U_0^2g_{\mu\rho}g^\dagger_{\sigma\nu}+
    \frac{1}{2}n_{imp}U_0^2\bm{g}_{\mu\rho}\bm{g}^\dagger_{\sigma\nu},
\end{equation}
where $\hat{\bm{g}}=i\hat{\bm{\sigma}}\hat\sigma_2$. We introduce
an impurity-renormalized gap function
$D_{\alpha\beta}(\bq,\omega_n)$, which satisfies an integral
equation
\begin{eqnarray}
\label{D eq}
    &&\hat D(\bq,\omega_n)=\eta(\bq)\hat g\nonumber\\
    &&\quad +\frac{1}{2}n_{imp}U_0^2\hat g
    \int\frac{d^3\bk}{(2\pi)^3}\tr\bigl[\hat g^\dagger\hat
    G(\bk+\bq,\omega_n)\nonumber\\
    &&\qquad \times\hat D(\bq,\omega_n)\hat G^T(-\bk,-\omega_n)\bigr]\nonumber\\
    &&\quad +\frac{1}{2}n_{imp}U_0^2\hat{\bm{g}}\int\frac{d^3\bk}{(2\pi)^3}\tr\bigl[\hat{\bm{g}}^\dagger
    \hat G(\bk+\bq,\omega_n)\nonumber\\
    &&\qquad \times\hat D(\bq,\omega_n)\hat G^T(-\bk,-\omega_n)\bigr].
\end{eqnarray}
Seeking solution in the form
\begin{equation}
\label{D st}
    \hat D(\bq,\omega_n)=d_0(\bq,\omega_n)\hat
    g+\bm{d}(\bq,\omega_n)\hat{\bm{g}},
\end{equation}
we obtain the following equations for $d_a(\br,\omega_n)$
($a=0,1,2,3$):
\begin{equation}
\label{dd eqs}
    d_a(\bq,\omega_n)=\eta(\bq)\delta_{a0}
    +\Gamma\sum_{b=0}^3\mathcal{Y}_{ab}(\bq,\omega_n)d_b(\bq,\omega_n),
\end{equation}
where $\eta(\bq)$ is the Fourier-transform of the order parameter,
and
\begin{eqnarray}
\label{Yab}
    &&\mathcal{Y}_{ab}(\bq,\omega_n)=\frac{1}{2\pi N_F}\nonumber\\
    &&\ \times\int\frac{d^3\bk}{(2\pi)^3}\tr\bigl[
    \hat{\mathrm{g}}_a^\dagger\hat G(\bk+\bq,\omega_n)
    \hat{\mathrm{g}}_b\hat G^T(-\bk,-\omega_n)\bigr],\qquad
\end{eqnarray}
with $\hat{\mathrm{g}}_0=\hat g$, and $\hat{\mathrm{g}}_i=\hat
g_i$ for $i=1,2,3$. We see that, in addition to the spin-singlet
component $d_0(\bq,\omega_n)$ of the gap function, impurity
scattering can induce a nonzero spin-triplet component
$\bm{d}(\bq,\omega_n)$. Note that the free energy depends only on
the singlet component: From Eqs. (\ref{dd eqs}) and (\ref{X def}),
we obtain:
\begin{eqnarray}
\label{F d0}
    {\cal F}&=&\int\frac{d^3\bq}{(2\pi)^3}\eta^*(\bq)\biggl[\frac{1}{V}
    \eta(\bq)\nonumber\\
    &&-\pi
    N_FT\sum_n{}'\frac{d_0(\bq,\omega_n)-\eta(\bq)}{\Gamma}\biggr].
\end{eqnarray}

\begin{figure}[t]
    \includegraphics[width=8.1cm]{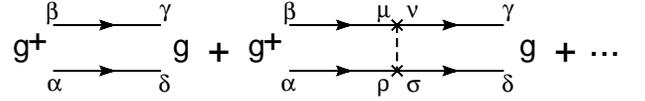}
    \caption{Impurity ladder diagrams in the Cooper channel. Lines with arrows
    correspond to the average Green's functions of electrons in the spin representation,
    $\hat g=i\hat\sigma_2$, and the impurity (dashed) lines
    are defined in the text, see Eq. (\ref{impurity line}).}
    \label{fig: ladder diagrams}
\end{figure}

Substituting the Green's functions (\ref{G clean zero H}) into
Eqs. (\ref{Yab}) and calculating the spin traces, we obtain for
the singlet-singlet part of the $4\times 4$ matrix
$\hat{\mathcal{Y}}$:
\begin{eqnarray}
\label{Y00}
    &&\mathcal{Y}_{00}(\bq,\omega_n)=\frac{1}{2\pi N_F}\nonumber\\
    &&\quad \times\sum_\lambda\int\frac{d^3\bk}{(2\pi)^3}
    G_\lambda(\bk+\bq,\omega_n)G_\lambda(-\bk,-\omega_n)\quad\nonumber\\
    &&\ =\frac{1}{2}\sum_\lambda\rho_\lambda
    \left\langle
    \frac{1}{|\omega_n|+\Gamma+i\Omega\,\sign\omega_n}
    \right\rangle_{\hat{\bk}}\nonumber\\
    &&\ =\left\langle
    \frac{1}{|\omega_n|+\Gamma+i\Omega\,\sign\omega_n}
    \right\rangle_{\hat{\bk}},\quad
\end{eqnarray}
where
\begin{equation}
\label{rhos}
    \rho_\pm=\frac{N_\pm}{N_F}=1\pm\delta
\end{equation}
are the fractional densities of states in the two bands,
$\Omega(\bk,\bq)=v_F\hat\bk\bq/2$, and
$\langle(...)\rangle_{\hat{\bk}}$ denotes the averaging over the
directions of $\bk$. The parameter
\begin{equation}
\label{delta def}
    \delta=\frac{N_+-N_-}{N_++N_-},
\end{equation}
which characterizes the difference between the band densities of
states, can be expressed in terms of the SO coupling strength
(\ref{SOC strength}) as follows:
\begin{equation}
\label{delta vs rho}
    |\delta|=\frac{2\varrho\sqrt{1+\varrho^2}}{1+2\varrho^2}.
\end{equation}
Similarly, for the singlet-triplet mixing part we obtain:
\begin{eqnarray}
\label{Y0i}
    &&\mathcal{Y}_{0i}(\bq,\omega_n)=\mathcal{Y}_{i0}(\bq,\omega_n)=\frac{1}{2\pi N_F}\nonumber\\
    &&\quad \times\sum_\lambda\lambda\int\frac{d^3\bk}{(2\pi)^3}\hat k_i
    G_\lambda(\bk+\bq,\omega_n)G_\lambda(-\bk,-\omega_n)\qquad\nonumber\\
    &&\ =\delta\left\langle
    \frac{\hat k_i}{|\omega_n|+\Gamma+i\Omega\,\sign\omega_n}
    \right\rangle_{\hat{\bk}}.
\end{eqnarray}
Finally, the triplet-triplet part can be represented as follows:
\begin{equation}
\label{Yij}
    \mathcal{Y}_{ij}(\bq,\omega_n)=\mathcal{Y}^{(1)}_{ij}(\bq,\omega_n)
    +\mathcal{Y}^{(2)}_{ij}(\bq,\omega_n),
\end{equation}
where
\begin{eqnarray*}
    &&\mathcal{Y}^{(1)}_{ij}(\bq,\omega_n)=\frac{1}{2\pi N_F}\\
    &&\ \times\sum_\lambda\int\frac{d^3\bk}{(2\pi)^3}
    \hat k_i\hat
    k_jG_\lambda(\bk+\bq,\omega_n)G_\lambda(-\bk,-\omega_n)\\
    &&\ =\left\langle
    \frac{\hat k_i\hat k_j}{|\omega_n|+\Gamma+i\Omega\,\sign\omega_n}
    \right\rangle_{\hat{\bk}},
\end{eqnarray*}
and
\begin{eqnarray*}
    &&\mathcal{Y}^{(2)}_{ij}(\bq,\omega_n)\\
    &&\ =\frac{1}{2\pi N_F}\sum_\lambda\int\frac{d^3\bk}{(2\pi)^3}
    (\delta_{ij}-\hat k_i\hat k_j-i\lambda e_{ijl}\hat k_l)\\
    &&\qquad\times
    G_\lambda(\bk+\bq,\omega_n)G_{-\lambda}(-\bk,-\omega_n)\\
    &&\ =\frac{1}{2}\sum_\lambda\left\langle
    \frac{\delta_{ij}-\hat k_i\hat k_j-i\lambda e_{ijl}\hat k_l}{|\omega_n|+\Gamma
    +i(\Omega+\lambda E_{SO}/2)\sign\omega_n}
    \right\rangle_{\hat{\bk}}.\quad
\end{eqnarray*}
We see that in the band representation the singlet impurity
scattering channel, which is described by the first term in Eq.
(\ref{impurity line}), causes only the scattering of intraband
pairs between the bands. In contrast, the triplet channel can mix
intraband and interband pairs, the latter being described by
$\mathcal{Y}^{(2)}_{ij}$.

The superconducting critical temperature is found by setting
$\bq=0$, then $\mathcal{Y}_{0i}=\mathcal{Y}_{i0}=0$, and the
solution of Eqs. (\ref{dd eqs}) has the form
$d_0=(1+\Gamma/|\omega_n|)\eta$. Substituting this into Eq.
(\ref{F d0}) and comparing the result with Eq. (\ref{F gen}), we
obtain:
\begin{equation}
\label{S uniform}
    S(\bq=0)=\frac{1}{V}-\pi N_FT\sum_n{}'\frac{1}{|\omega_n|},
\end{equation}
which yields the superconducting critical temperature:
\begin{equation}
\label{Tc0}
    T_{c0}=\frac{2e^{\mathbb{C}}}{\pi}\omega_c e^{-1/N_FV},
\end{equation}
where $\mathbb{C}\simeq 0.577$ is Euler's constant. Thus there is
an analog of Anderson's theorem in noncentrosymmetric
superconductors with a BCS-contact pairing interaction: The
zero-field critical temperature is not affected by scalar
disorder.\cite{MS07}

Now let us turn on a uniform magnetic field $\bH=H\hat z$. Its
orbital effect is taken into account in the usual way, by
replacing $\bq\to\bD=-i\bnab+(2e/c)\bm{A}$ in
$\mathcal{Y}_{ab}(\bq,\omega_n)$, which become differential
operators of infinite order. The gap equations (\ref{dd eqs}) then
take the following form:
\begin{equation}
\label{gap eqs}
    \left\{\begin{array}{l}
      (1-\Gamma\hat{\mathcal{Y}}_{00})d_0-\Gamma\hat{\mathcal{Y}}_{0i}d_i=\eta,
      \medskip\\
      -\Gamma\hat{\mathcal{Y}}_{i0}d_0+(\delta_{ij}-\Gamma\hat{\mathcal{Y}}_{ij})d_j=0.  \\
    \end{array}\right.
\end{equation}
Solution of these equations in the general case is rather
cumbersome. In order to make progress, we use the fact that in
practice the SO coupling is much weaker than the Fermi energy. In
terms of the parameter $\delta$, this translates into the
following condition: $|\delta|\ll 1$. This allows us to neglect
the triplet component of the gap function. Indeed, using the fact
that the singlet-triplet mixing is described by
$\mathcal{Y}_{0i}=\mathcal{Y}_{i0}$, which are proportional to
$\delta$, see Eq. (\ref{Y0i}), one can solve the gap equations
(\ref{gap eqs}) by iterations. It is easy to see that
$\bm{d}\sim\mathcal{O}(\delta)$, therefore the correction to
$d_0=(1-\Gamma\hat{\mathcal{Y}}_{00})^{-1}\eta$ due to the
impurity-induced singlet-triplet mixing is of the order of
$\delta^2$ and will be neglected.

\subsection{Equation for $H_{c2}(T)$}
\label{sec: Hc2 equation}

Keeping only the singlet gap component, we have
$d_0=(1-\Gamma\hat{\mathcal{Y}}_{00})^{-1}\eta$. Substituting this
in Eq. (\ref{F d0}), we can represent the operator $\hat S$ as
follows:
\begin{equation}
\label{S vs Y00}
    \hat S=\frac{1}{V}-\pi N_FT\sum_n{}'[1-\Gamma\hat{\mathcal{Y}}_{00}(\omega_n)]^{-1}
    \hat{\mathcal{Y}}_{00}(\omega_n),
\end{equation}
where $\hat{\mathcal{Y}}_{00}(\omega_n)$ is defined by the kernel
\begin{eqnarray}
\label{X_0}
    &&\mathcal{Y}_{00}(\br_1,\br_2;\omega_n)=\frac{1}{2\pi N_F}
    g^\dagger_{\alpha\beta}g_{\gamma\delta}\nonumber\\
    &&\quad\times\bigl\langle G_{\beta\gamma}(\br_1,\br_2;\omega_n)\bigr\rangle_{imp}
    \bigl\langle G_{\alpha\delta}(\br_1,\br_2;-\omega_n)\bigr\rangle_{imp}.\quad
\end{eqnarray}
Although the expression (\ref{S vs Y00}) is approximate (with the
corrections of the order of $\delta^2$), it has the advantage of
being relatively simple and captures important physics of the
problem, including the properties of various nonuniform
superconducting states created by the external field, see Secs.
\ref{sec: GL regime}-\ref{sec: General case} below. The general
case of arbitrary SO coupling, with both singlet and triplet
channels present, will be considered elsewhere.\cite{Sam08}

It is shown in Appendix A that the eigenfunctions of
$\hat{\mathcal{Y}}_{00}(\omega_n)$ are the Landau levels
$|N,p\rangle$, which are labelled by two quantum numbers: a
non-negative integer $N$ and a real $p$. The latter is
proportional to the wave vector of the order parameter modulation
along the applied field: $p=\ell_Hq_z$, where $\ell_H=\sqrt{c/eH}$
is the magnetic length. The corresponding eigenvalues can be
written as follows:
\begin{equation}
\label{y def}
    \hat{\mathcal{Y}}_{00}(\omega_n)|N,p\rangle=y_{N,p}(\omega_n)|N,p\rangle,
\end{equation}
where
\begin{eqnarray}
\label{y N p}
    &&y_{N,p}(\omega_n)=\int_0^\infty
    du\;e^{-(|\omega_n|+\Gamma)u}\nonumber\\
    &&\qquad\times\int_0^1 ds\,
    F_p(u,s)e^{-v^2(1-s^2)/2}L_N[v^2(1-s^2)].\qquad
\end{eqnarray}
Here $v=(v_F/2\ell_H)u$, $L_N(x)$ are the Laguerre polynomials,
\begin{eqnarray}
\label{Fus}
    F_p(u,s)=\cos(p_0vs)\cos(pvs)-\delta\sin(p_0vs)\sin(pvs),
\end{eqnarray}
$p_0=g\mu_BH\ell_H/v_F$, and $\delta$ is defined by Eq.
(\ref{delta def}). Note that in \emph{centrosymmetric} isotropic
superconductors with nonmagnetic impurities, the eigenvalues of
$\hat{\mathcal{Y}}_{00}(\omega_n)$ are given by the same
expression (\ref{y N p}), but with $F_p(u,s)=\cos(p_0v)\cos(pvs)$.
This is still different from Eq. (\ref{Fus}), even if one sets
$\delta=0$ in the latter. The reason is that Eq. (\ref{Fus}) is
valid under the assumption that the Zeeman energy is small
compared with the SO band splitting.

Since the Landau levels are eigenfunctions of
$\hat{\mathcal{Y}}_{00}(\omega_n)$ at all frequencies, the
operator $\hat S$, see Eq. (\ref{S vs Y00}), is also diagonal in
the basis of $|N,p\rangle$. Using Eqs. (\ref{S uniform}) and
(\ref{Tc0}), we can eliminate both the frequency cutoff and the
coupling constant from Eq. (\ref{S vs Y00}). In this way we obtain
an equation for the magnetic field at which a superconducting
instability characterized by $N$ and $p$ develops:
\begin{equation}
\label{Hc2 eq}
    \ln\frac{T_{c0}}{T}=\pi T\sum_n\left[\frac{1}{|\omega_n|}-
    \frac{y_{N,p}(\omega_n)}{1-\Gamma y_{N,p}(\omega_n)}\right].
\end{equation}
The upper critical field, $H_{c2}(T)$, is obtained by maximizing
the solution of the above equation with respect to both $N$ and
$p$.

It is convenient to introduce the reduced temperature, magnetic
field, and disorder as follows:
\begin{equation}
\label{reduced parameters}
    t=\frac{T}{T_{c0}},\quad h=\frac{2H}{H_0},\quad
    \zeta=\frac{\Gamma}{\pi T_{c0}},
\end{equation}
where $H_0=\Phi_0/\pi\xi_0^2$, $\Phi_0=\pi c/e$ is the magnetic
flux quantum, and $\xi_0=v_F/2\pi T_{c0}$ is the superconducting
coherence length. Eq. (\ref{Hc2 eq}) then takes the form
\begin{equation}
\label{Hc2 eq reduced}
    \ln\frac{1}{t}=2\sum_{n\geq 0}\left[\frac{1}{2n+1}-t
    \frac{w_n^{(N,Q)}(t,h)}{1-\zeta w_n^{(N,Q)}(t,h)}\right],
\end{equation}
where
\begin{eqnarray}
\label{w N Q}
    w_n^{(N,Q)}(t,h)=\int_0^\infty d\rho\,e^{-[(2n+1)t+\zeta]\rho}\int_0^1
    ds\,\Phi_Q(\rho,s)\nonumber\\
    \times
    \exp\left[-\frac{h}{4}\rho^2(1-s^2)\right]L_N\left[\frac{h}{2}\rho^2(1-s^2)\right],
\end{eqnarray}
and
\begin{eqnarray}
\label{Phi Q}
    \Phi_Q(\rho,s)&=&\cos(\alpha h\rho s)\cos(Q\rho s)\nonumber\\
    &&\qquad-\delta\sin(\alpha h\rho s)\sin(Q\rho s).\quad
\end{eqnarray}
The parameter
\begin{equation}
\label{alpha def}
    \alpha=\frac{g}{2}\frac{\mu_BH_0}{2\pi T_{c0}}
\end{equation}
measures the relative importance of the paramagnetic and orbital
contributions to the magnetic suppression of superconductivity
(note that $\alpha$ is proportional to the Maki parameter
$\alpha_M$, introduced in Ref. \onlinecite{Maki64}). For the upper
critical field in the reduced notations we have
\begin{equation}
    h_{c2}(t)=\max_{N,Q}h_{N,Q}(t),
\end{equation}
where $h_{N,Q}(t)$ is the solution of Eq. (\ref{Hc2 eq reduced}),
$N$ is the Landau level index, and $Q=\xi_0q_z$ is the
dimensionless wave vector of the superconducting instability.

The purely orbital limit is obtained by formally setting $g$ to
zero. Then $\alpha=0$, and Eq. (\ref{Hc2 eq reduced}) takes
exactly the same form as in centrosymmetric BCS superconductors.
Therefore, if the Zeeman interaction is neglected then the absence
of inversion symmetry does not bring about any new features in
$H_{c2}(T)$, compared with the centrosymmetric case, described by
the Helfand-Werthamer theory.\cite{HW66} The maximum critical
field corresponds to $N=Q=0$ at all temperatures. In particular,
at $T=0$ and in the absence of impurities one obtains:
$H_{c2}(0)=(e^{2-\mathbb{C}}/8)H_0\simeq 0.52H_0$. This can be
used to relate $\alpha$ to experimentally observable quantities as
follows:
\begin{equation}
\label{alpha exp}
    \alpha\simeq 0.21\frac{H_{c2}(0)[T]}{T_{c0}[K]}.
\end{equation}
Here we assumed $g=2$.

\section{Ginzburg-Landau regime}
\label{sec: GL regime}

At weak external field, the critical temperature is close to
$T_{c0}$, i.e. the reduced temperature $t$ is close to $1$. In
this limit, one can solve Eq. (\ref{Hc2 eq reduced}) analytically,
by expanding $w_n^{(N,Q)}$ in powers of $h$ and $Q$ (we set
$N=0$). We seek solution in the form
\begin{equation}
\label{tc expansion}
    t_c(h)=1-a_1h-a_2h^2+\mathcal{O}(h^3).
\end{equation}
It follows from Eq. (\ref{Phi Q}) that the maximum critical
temperature corresponds to
\begin{equation}
\label{Q maximum}
    Q=-\delta\alpha h.
\end{equation}
After some straightforward algebra, we obtain the following
expressions for the coefficients in the expansion (\ref{tc
expansion}):
\begin{eqnarray}
\label{a1}
    a_1&=&\frac{1}{3}\mathcal{S}_{2,1},\\
\label{a2}
    a_2&=&\frac{1}{18}\mathcal{S}_{2,1}^2+\frac{1}{9}\mathcal{S}_{2,1}\mathcal{S}_{1,2}
    -\frac{2}{5}\mathcal{S}_{2,3}-\frac{1}{9}\zeta\mathcal{S}_{3,3}\nonumber\\
    &&+\frac{1}{3}\alpha^2\mathcal{S}_{2,1},
\end{eqnarray}
with
$$
    \mathcal{S}_{k,l}(\zeta)=2\sum_{n\geq 0}\frac{1}{(2n+1)^k(2n+1+\zeta)^l}.
$$
Since our model is valid only in the limit of weak SO coupling,
$|\delta|\ll 1$, we have omitted in these expressions the terms
containing $\delta^2$. In the clean limit, we have
\begin{eqnarray*}
    && a_1=\frac{7\zeta(3)}{12},\\
    && a_2=\frac{49\zeta^2(3)}{96}-\frac{31\zeta(5)}{40}
    +\frac{7\zeta(3)}{12}\alpha^2,
\end{eqnarray*}
where $\zeta(x)$ is the Riemann zeta function.

Since Eq. (\ref{a1}) does not contain $\alpha$, the slope of the
upper critical field near $T_{c0}$  is entirely determined by the
orbital effects. Returning to dimensional variables, we obtain:
\begin{equation}
\label{Hc2 GL orbital}
    \left|\frac{dH_{c2}}{dT}\right|_{T=T_{c0}}
    =\frac{6\pi}{\mathcal{S}_{2,1}(\zeta)}\frac{\Phi_0T_{c0}}{v_F^2},
\end{equation}
which has exactly the same form as in isotropic centrosymmetric
superconductors.\cite{Gor60} From this expression it follows, in
particular, that the upper critical field slope in our model is
enhanced by disorder, in agreement with the results obtained in
Ref. \onlinecite{MS07}.

The effects of the Zeeman interaction on the critical temperature
appear only in the second order in $h$ and are described by the
last term in Eq. (\ref{a2}). If the difference between the band
densities of states is taken into account (i.e. if $\delta\neq
0$), then the order parameter is modulated along the applied
field, with the period given by Eq. (\ref{Q maximum}).

In the limit of large $\alpha$, the Zeeman term dominates and the
critical field slope diverges: $H_{c2}(T)\sim\sqrt{T_{c0}-T}$.
Similar to the orbital critical field (\ref{Hc2 GL orbital}), the
paramagnetic critical field in the Ginzburg-Landau regime is
enhanced by disorder. The paramagnetic pair breaking in
noncentrosymmetric superconductors has some peculiar features,
compared with the centrosymmetric case, and is studied in detail
in the next section.

\section{Paramagnetic limit}
\label{sec: Paramagnetic}

The purely paramagnetic limit corresponds to $\alpha\to\infty$ in
Eq. (\ref{Hc2 eq reduced}). Due to the fast oscillations of
$\Phi_Q$, the last two factors in the $s$-integral in Eq. (\ref{w
N Q}) can be replaced by 1, and the solution becomes degenerate
with respect to $N$. We have
\begin{eqnarray}
\label{w Q paramag}
    &&w_n(Q)=\frac{1}{2}\sum_\lambda\rho_\lambda\nonumber\\
    &&\qquad\times\re\left\langle
    \frac{1}{(2n+1)t+\zeta+i(Q+\lambda\tilde h)\hat
    k_z}\right\rangle_{\hat\bk},
\end{eqnarray}
where
\begin{equation}
\label{tilde h def}
    \tilde h=\frac{g}{2}\frac{\mu_BH}{\pi T_{c0}}=\alpha h.
\end{equation}
Calculating the Fermi-surface integrals, we obtain:
\begin{eqnarray}
\label{w Q paramag final}
    w_n(Q)=\frac{1}{2}\sum_\lambda\rho_\lambda\frac{1}{Q+\lambda\tilde h}
    \arctan\frac{Q+\lambda\tilde h}{(2n+1)t+\zeta}.
\end{eqnarray}
Further analytical progress can be made in two limiting cases:
clean ($\zeta=0$) and ``dirty'' ($\zeta\gg 1$), while the
intermediate disorder strengths can only be studied numerically.

\subsection{Clean case}
\label{sec: para clean}

In the clean limit, it is convenient to go back to expression
(\ref{w Q paramag}) and substitute it into Eq. (\ref{Hc2 eq
reduced}), with the following result:
\begin{eqnarray}
\label{Hc para eq}
    \ln\frac{1}{t}=\frac{1}{2}\sum_\lambda\rho_\lambda\left\langle\re\Psi\left(\frac{1}{2}+
    i\frac{Q+\lambda\tilde h}{2t}\hat k_z\right)\right\rangle_{\hat\bk}\nonumber\\
    -\Psi\left(\frac{1}{2}\right),
\end{eqnarray}
where $\Psi(x)$ is the digamma function. From this we obtain:
\begin{equation}
\label{tc paramag}
    t=\exp\Bigl[\max_y\mathcal{I}(y,z)\Bigr],
\end{equation}
where $y=Q/2t$, $z=\tilde h/2t$,
\begin{eqnarray*}
    \mathcal{I}(y,z)=\Psi\left(\frac{1}{2}\right)
    -\frac{1}{2}\sum_\lambda\rho_\lambda
    \frac{\im\ln\Gamma(1/2+iy+i\lambda z)}{y+\lambda z},
\end{eqnarray*}
and $\Gamma(x)$ is the Gamma function. At any given $z$, if the
maximum of $\mathcal{I}(y,z)$ is achieved at $y=y_c$, then the
wave vector of the superconducting instability is $Q=2ty_c$, and
the corresponding critical field is $\tilde h=2tz$. The critical
field of a second-order phase transition into a uniform
superconducting state can be found by setting $y=0$ in Eq.
(\ref{tc paramag}). In particular, at zero temperature this phase
transition occurs at $\tilde h_0=1/2e^{\mathbb{C}-1}\simeq 0.76$.

The pair-breaking effect of the Zeeman interaction can be reduced
by allowing the pairs to have a nonzero center-of-mass momentum.
The outcome of the competition between the Zeeman energy and the
gradient energy depends on the difference between the densities of
states in the two bands, which in turn depends on the ratio of the
SO band splitting to the Fermi energy, see Eqs. (\ref{rhos}) and
(\ref{delta vs rho}). If it is neglected, i.e. $\delta=0$, then
one obtains from Eq. (\ref{tc paramag}) that at $z<z^*\simeq
0.44$, which corresponds to high temperatures, $t>t^*\simeq 0.68$,
and low fields, $\tilde h<\tilde h^*\simeq 0.60$, the maximum
critical field is achieved at $Q=0$, and the phase transition
occurs into the uniform superconducting state. However, at $t<t^*$
the maximum critical field corresponds to $Q\neq 0$, and the phase
transition occurs into a nonuniform superconducting state, similar
to the Larkin-Ovchinnikov-Fulde-Ferrell (LOFF) state.\cite{LOFF}
The order parameter is modulated along the field:
$\eta(\br)=\eta_1e^{iqz}+\eta_2e^{-iqz}$, where $q=Q/\xi_0$. The
coefficients $\eta_{1,2}$ are found from the higher-order terms in
the free energy. In Refs. \onlinecite{DF07} and \onlinecite{AK07},
this was done for a Rashba superconductor, and it was shown that
both a multiple-$q$ (or stripe) LOFF state and a single plane wave
state are possible, depending on temperature.

\begin{figure}[t]
    \includegraphics[width=8.2cm]{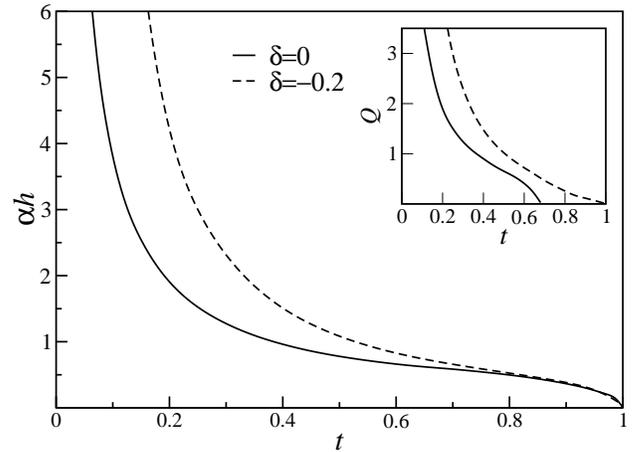}
    \caption{Paramagnetic critical field $\tilde h=\alpha h$
    in the clean case, for the LOFF state ($\delta=0$) and the
    helical state with $\delta=-0.2$.
    Inset shows the temperature dependence of the wave vector $Q$
    of the order parameter modulation along the field.}
    \label{fig: hc para clean}
\end{figure}

If the difference between $\rho_+$ and $\rho_-$ is taken into
account, then the right-hand side of Eq. (\ref{Hc para eq}) is no
longer even in $Q$, and the maximum critical field corresponds to
$Q\neq 0$ at all temperatures $0\leq t<1$. In this case, the phase
transition occurs into a single plane wave, or
helical,\cite{Agter03} superconducting state with
$\eta(\br)=\eta_0e^{iqz}$. In particular, at weak fields near
$t=1$, the maximum of $\mathcal{I}(y,z)$ is achieved at
$y_c=-\delta z$, which corresponds to $Q=-\delta\tilde h$. For the
critical field one obtains:
\begin{equation}
\label{tc paramag low h}
    \tilde h|_{t\to 1}=\sqrt{\frac{12}{7\zeta(3)}}(1-t)^{1/2}.
\end{equation}
At low temperatures, we use the fact that
$\max_y\mathcal{I}(y,z)=-(\rho_+/2)\ln(8e^{\mathbb{C}-1}z)$ in the
limit $z\to\infty$ (for $\rho_+\leq\rho_-$). Therefore,
\begin{equation}
\label{tilde h low t}
    \tilde h|_{t\to 0}=\frac{1}{4e^{\mathbb{C}-1}}t^{-\rho_-/\rho_+}.
\end{equation}
In these expressions we have omitted the terms proportional to
$\delta^2$. The temperature dependence of the critical field is
shown in Fig. \ref{fig: hc para clean}, both in the LOFF state,
for $\rho_+=\rho_-=1$, and in the helical state, for $\rho_+=0.8$,
$\rho_-=1.2$.

The most notable feature of the phase diagram is a considerable
weakening of the paramagnetic pair breaking at low temperatures,
which is manifested in the divergence of the critical field at
$t\to 0$.\cite{divergence} Similar behavior has also been found in
two-dimensional noncentrosymmetric
superconductors.\cite{BG02,DF07}

We would like to mention that the paramagnetic pair breaking
disappears altogether in the (rather unrealistic) extreme
single-band case $|\delta|=1$, which corresponds to a very strong
SO coupling, $\varrho\to\infty$, see Eq. (\ref{delta vs rho}).
Although a quantitative treatment of this case is beyond the
limits of applicability of our model, one can use simple arguments
to show that the critical temperature is not affected by the
applied field. We first note that the Zeeman interaction causes an
anisotropic deformation of the electron bands:
$\xi_\lambda(\bk)\to\xi_\lambda(\bk)+\lambda(g/2)\mu_B\hat\bk\bH$.
The intraband pairing of electrons with opposite momenta costs
additional energy at $H\neq 0$, resulting in the uniform
superconductivity being suppressed by the Zeeman field. On the
other hand, it is easy to see that the field-induced deformation
of the bands amounts to shifting the bands in the opposite
directions along the field:
$\xi_\lambda(\bk)\to\xi_\lambda(\bk+\bq_\lambda)$, where
$\bq_\lambda=\lambda(g/2)\mu_B\bH/v_F$. If the Cooper pairs in the
``$+$'' and ``$-$'' bands are completely decoupled, or if there is
just one band present, then the band shifts can be eliminated by
independent gauge transformations, and the Zeeman pair breaking
will be absent. According to Ref. \onlinecite{SM08}, in the
BCS-contact model (\ref{H_int}) the pairs in the two bands are in
fact strongly coupled, so that both condensates are characterized
by the same wave function $\eta(\br)$. The band shifts cannot be
eliminated simultaneously in both bands by any gauge
transformation. Therefore, in general there is some energy penalty
associated with the Cooper pairing (both uniform and nonuniform)
at $H\neq 0$, compared with the zero-field case.

\subsection{Disordered case}
\label{sec: para dis}

\begin{figure}[t]
    \includegraphics[width=8.2cm]{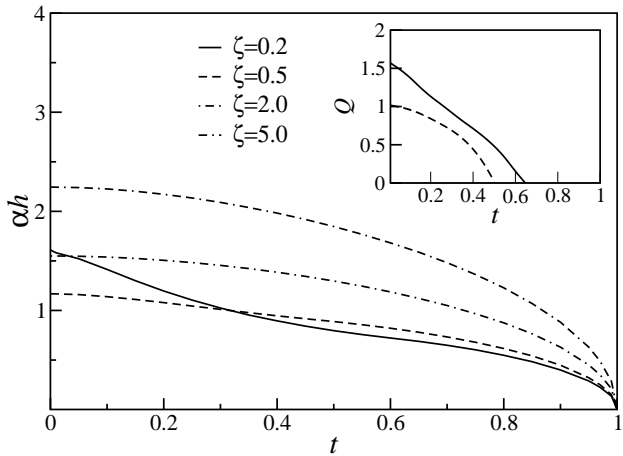}
    \caption{Paramagnetic critical field in the LOFF state ($\delta=0$)
    for different strengths of disorder. Inset shows the order parameter
    modulation along the field
    (for $\zeta=2.0$ and $5.0$, $Q=0$ at all temperatures).}
    \label{fig: hc para dis LOFF}
\end{figure}

\begin{figure}[t]
    \includegraphics[width=8.2cm]{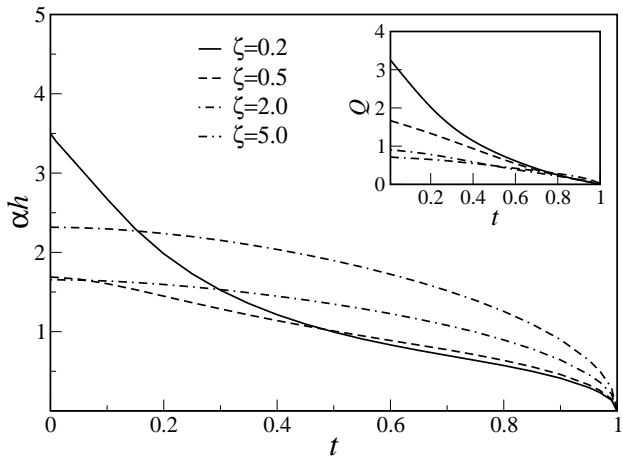}
    \caption{Paramagnetic critical field in the helical state with
    $\delta=-0.2$, for different strengths of disorder. Inset shows
    the order parameter modulation along the field.}
    \label{fig: hc para dis helical}
\end{figure}

At arbitrary disorder, the paramagnetic critical field is obtained
by numerical solution of Eq. (\ref{Hc2 eq reduced}), with $w_n$
given by expressions (\ref{w Q paramag final}). At $\delta=0$, we
find that the LOFF modulation is suppressed by impurities, see
Fig. \ref{fig: hc para dis LOFF}, and disappears at
$\zeta>\zeta_c\simeq 1.16$. In contrast, the helical modulation at
$\delta\neq 0$ survives even if the impurity scattering is strong,
as shown in Fig. \ref{fig: hc para dis helical}. In both cases,
the low-temperature divergence of the critical field
characteristic of a clean system, see Eq. (\ref{tilde h low t}),
is removed by disorder.

In the ``dirty'' limit $\zeta\gg 1$, one can obtain an equation
for the critical field in a closed form. We shall see that both
$Q$ and $\tilde h$ scale as $\sqrt{\zeta}\ll\zeta$. Therefore, one
can perform the Taylor expansion in Eq. (\ref{w Q paramag final}):
$$
    w_n(Q)\simeq\frac{1}{(2n+1)t+\zeta}\left[1-\frac{W}{((2n+1)t+\zeta)^2}\right],
$$
where $W=(Q^2+\tilde h^2+2\delta Q\tilde h)/3$. The main
contribution to the Matsubara sum in Eq. (\ref{Hc2 eq reduced})
comes from $(2n+1)t\ll\zeta$, and we obtain:
$\ln(1/t)=\Psi(1/2+W/2\zeta t)-\Psi(1/2)$. It is easy to see that
the maximum critical temperature is achieved when $W$ has a
minimum with respect to $Q$, which happens for $Q=-\delta\tilde
h$, at all temperatures. Thus we arrive at a well-known universal
equation, which describes magnetic pair breaking in
superconductors:\cite{Tink-book}
\begin{equation}
\label{universal eq}
    \ln\frac{1}{t}=\Psi\left(\frac{1}{2}+\frac{\sigma}{t}\right)
    -\Psi\left(\frac{1}{2}\right).
\end{equation}
The pair-breaker strength is characterized by $\sigma=\tilde
h^2/6\zeta$. Analytical expressions for the critical field can be
obtained in the weak-field limit:
\begin{equation}
\label{tc paramag dis low h}
    \tilde h|_{t\to 1}=\sqrt{\frac{12\zeta}{\pi^2}}(1-t)^{1/2},
\end{equation}
and also at low temperatures:
\begin{equation}
\label{tilde h dis low t}
    \tilde h|_{t\to 0}=\sqrt{\frac{3\zeta}{2e^{\mathbb{C}}}}.
\end{equation}

Thus we come to the conclusion that, while the critical field near
$T_{c0}$ is enhanced by disorder, the impurity response at low
temperatures is non-monotonic: Initially, the disorder cuts off
the singularity at $T\to 0$, thus reducing the critical field.
However, as the disorder increases one eventually reaches the
universal regime, in which the critical field grows as
$\sqrt{\zeta}$.

\subsection{First-order phase transition}
\label{sec: FOPT}

Let us now compare the paramagnetic critical fields found above
with the critical field of a first-order phase transition (FOPT)
into a uniform superconducting state at zero temperature. For
simplicity, we neglect the difference between the band densities
of states and set $g=2$. Following the arguments of Clogston and
Chandrasekhar, \cite{CC62} we obtain:
$\mu_BH_{FOPT}=(\eta_0/\sqrt{2})(1-\chi_s/\chi_P)^{-1/2}$, where
$\eta_0=(\pi/e^{\mathbb{C}})T_{c0}$ is the gap magnitude at zero
temperature and field, $\chi_s$ is the residual susceptibility at
$T=0$ in the superconducting state, and $\chi_P=2\mu_B^2N_F$ is
the Pauli susceptibility in the normal state. Therefore,
\begin{equation}
\label{FOPT}
    \tilde h_{FOPT}|_{t\to 0}=\frac{1}{\sqrt{2}\,e^{\mathbb{C}}}
    \left(1-\frac{\chi_s}{\chi_P}\right)^{-1/2}.
\end{equation}
According to Ref. \onlinecite{Sam07}, the residual susceptibility
in our model is given by the following expression:
\begin{equation}
    \frac{\chi_s}{\chi_P}=\frac{2}{3}+\frac{1}{3}\Phi(x),
\end{equation}
where $x=2\Gamma/3\eta_0$, and
$$
    \Phi(x)=1-\frac{\pi}{2x}\biggl(1-\frac{4}{\pi\sqrt{1-x^2}}
    \arctan\sqrt{\frac{1-x}{1+x}}\biggr)
$$
[at $x>1$ this function is evaluated using
$\arctan(ix)=i\tanh^{-1}(x)$]. In the clean limit, $\Phi(0)=0$ and
$\chi_s/\chi_P=2/3$. Therefore, $\tilde
h_{FOPT}=\sqrt{3/2}\,e^{-\mathbb{C}}\simeq 0.69$, which is much
lower than the divergent expression (\ref{tilde h low t}). In the
``dirty'' limit $x\to\infty$, we have $\Phi(x)\simeq 1-\pi/2x$,
and it follows from Eq. (\ref{FOPT}) that $\tilde
h_{FOPT}=\sqrt{2\zeta/\pi e^{\mathbb{C}}}$. Although the critical
field of the first-order phase transition into a uniform
superconducting state increases with disorder, it remains lower
than the critical field of the second-order phase transition into
the helical state, which is given by the expression (\ref{tilde h
dis low t}).

\section{General case: Orbital effects}
\label{sec: General case}

Solution of Eq. (\ref{Hc2 eq reduced}) at all temperatures and for
arbitrary values of $\alpha$ and $\zeta$ can only be obtained
numerically. To facilitate numerical analysis, we represent the
integrals in Eq. (\ref{w N Q}) in a somewhat shorter form by
introducing the Cartesian coordinates
$\bm{\rho}=(\rho_1,\rho_2,\rho_3)$, such that $\rho=|\bm{\rho}|$,
$\rho s=\rho_3$, and
$\rho^2(1-s^2)=\rho_\perp^2=\rho_1^2+\rho_2^2$. Using the Fourier
transform
$$
    \int d^3\bm{\rho}\,e^{i\bk\bm{\rho}}\frac{e^{-a\rho}}{\rho^2}=\frac{4\pi}{k}
    \arctan\frac{k}{a},
$$
we obtain:
\begin{eqnarray}
\label{w N Q final}
    w_n^{(N,Q)}(t,h)=\frac{1}{2}\sum_\lambda\rho_\lambda\int_0^\infty
    \frac{k\,dk}{\sqrt{k^2+(Q+\lambda\alpha h)^2}}\nonumber\\
    \times\arctan\biggl[\frac{\sqrt{k^2+(Q+\lambda\alpha
    h)^2}}{(2n+1)t+\zeta}\biggr]I_N(k),
\end{eqnarray}
where
\begin{equation}
    I_N=\frac{1}{2\pi}\int d^2\bm{\rho}_\perp\, e^{-i\bk\bm{\rho}_\perp}e^{-h\rho_\perp^2/4}
    L_N\Bigl(\frac{h\rho_\perp^2}{2}\Bigr).
\end{equation}
In the purely paramagnetic limit, $I_N=2\pi\delta(\bk)$, and one
recovers Eq. (\ref{w Q paramag final}).

The maximum critical field corresponds to $N=0$, in which case
$I_0(k)=(2/h)\exp(-k^2/h)$. The order parameter is nonuniform
along the direction of the applied field, while its dependence on
the transverse coordinates is given by the usual Abrikosov vortex
solution. As in the purely paramagnetic limit, the phase diagram
turns out to be different for $\delta=0$ (LOFF vortex state) and
$\delta\neq 0$ (helical vortex state).

Solution of Eqs. (\ref{Hc2 eq reduced}) and (\ref{w N Q final})
shows that the LOFF state is suppressed by the orbital effects,
even in the clean case. There is a critical strength of the
paramagnetic interaction, $\alpha_c$, below which the LOFF state
disappears, i.e. $Q=0$ at all temperatures (in the clean limit
$\alpha_c\simeq 0.66$). This is qualitatively similar to the way
the orbital interaction affects the LOFF state in centrosymmetric
superconductors, see Ref. \onlinecite{GG66}. In contrast, the
helical modulation at $\delta\neq 0$, albeit weakened by the
orbital effects, does not completely disappear until $\alpha=0$.

\begin{figure}[t]
    \includegraphics[width=8.2cm]{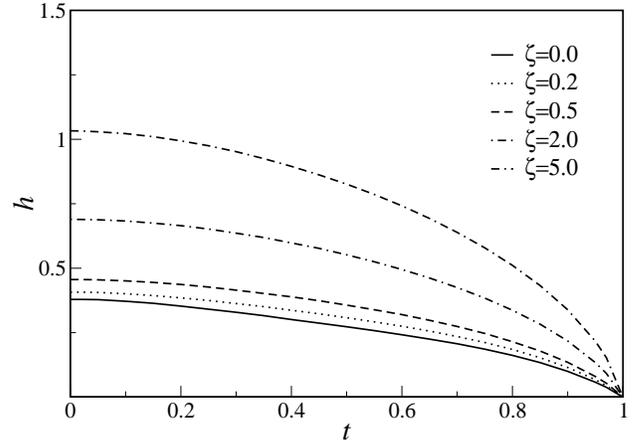}
    \caption{Upper critical field $h_{c2}(t)$ in the helical vortex state with $\alpha=2.0$ and
    $\delta=-0.2$, for different strengths of disorder.}
    \label{fig: h vs t vortex helical alpha 20}
\end{figure}

In order to study the combined effect of the orbital interaction
and impurities, we focus on the helical vortex state with
$\delta=-0.2$. We consider $\alpha=0.2$ [which is close to the
estimates for the Li$_2$(Pd$_{1-x}$,Pt$_x$)$_3$B family of
superconductors, as obtained from Eq. (\ref{alpha exp}) using the
data from Ref. \onlinecite{Hafl07}], and also $\alpha=2.0$. The
upper critical field curves for a range of disorder strengths are
shown in Figs. \ref{fig: h vs t vortex helical alpha 02} and
\ref{fig: h vs t vortex helical alpha 20}. We see that if $\alpha$
is not too large, disorder produces a monotonic enhancement of
$H_{c2}$ at all temperatures.

\begin{figure}[t]
    \includegraphics[width=8.2cm]{Fig5.eps}
    \caption{Upper critical field $h_{c2}(t)$ in the helical vortex state with $\alpha=0.2$ and
    $\delta=-0.2$, for different strengths of disorder.}
    \label{fig: h vs t vortex helical alpha 02}
\end{figure}

In the ``dirty'' limit, it is again possible to obtain a
relatively simple equation for the upper critical field. Repeating
the arguments from Sec. \ref{sec: para dis}, one can show that
that equation has the universal form (\ref{universal eq}), with
the paramagnetic and orbital interactions both contributing to the
pair-breaking strength:
\begin{equation}
\label{univeral pair breaking orbital}
    \sigma=\frac{h+\alpha^2h^2}{6\zeta}.
\end{equation}
The maximum critical field is achieved for $Q=-\delta\alpha h$.
Near $T_{c0}$, superconductivity is suppressed by magnetic field
according to Eq. (\ref{tc expansion}), in which
$a_1=\pi^2/12\zeta+O(\zeta^{-2})$ and
$a_2=(\pi^2/12\zeta)\alpha^2+O(\zeta^{-2})$. At low temperatures,
returning to dimensional variables, we arrive at the following
expression for the upper critical field:
\begin{equation}
\label{h dirty low t}
    H_{c2}(0)=\frac{\Phi_0}{2\pi\xi_0^2}\frac{1}{2\alpha^2}
    \Biggl(\sqrt{1+\frac{6\alpha^2}{\pi e^{\mathbb{C}}}
    \frac{\Gamma}{T_{c0}}}-1\Biggr).
\end{equation}
In the weak SO coupling limit, corrections to this expression are
of the order of $\delta^2$. In the orbital limit $\alpha\to 0$,
one obtains $H_{c2}(0)=(3/2\pi e^{\mathbb{C}})(\Gamma/T_{c0})$,
while in the paramagnetic limit $\alpha\to\infty$, Eq. (\ref{tilde
h dis low t}) is recovered. We see that the upper critical field
is enhanced by disorder.

\begin{figure}[t]
    \includegraphics[width=8.2cm]{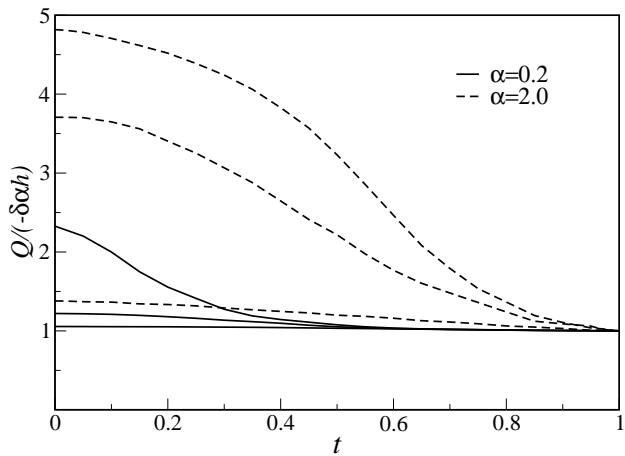}
    \caption{Ratio $Q/(-\delta\alpha h)$ in the helical vortex state
    with $\delta=-0.2$, for the following disorder strengths: $\zeta=0.0$, $0.5$,
    and $5.0$ (from top to bottom).}
    \label{fig: Q ratio}
\end{figure}

As for the wave vector of the helical modulation, the expression
(\ref{Q maximum}) is exact only in the Ginzburg-Landau regime and
in the ``dirty'' limit. In clean systems at low temperatures, deviations of
$Q/(-\delta\alpha h)$ from unity become quite substantial, see
Fig. \ref{fig: Q ratio}.

\section{Conclusions}
\label{sec: Conclusions}

We have calculated the upper critical field in noncentrosymmetric
superconductors at arbitrary temperatures, both in the clean case
and in the presence of scalar impurities, using as an example a
crystal of cubic symmetry with a BCS-contact pairing interaction.
Our derivation relies on the assumption that the SO band splitting
is small compared to the Fermi energy. Then one can neglect the
disorder-induced triplet component of the gap function, which
considerably simplifies the calculations, without losing
information about the interesting physics of the problem, in
particular about the properties of nonuniform superconducting
states.

As in the centrosymmetric case, the pair breaking is provided by
both the orbital and the Zeeman (paramagnetic) interactions. If
the former is dominant, i.e. $\alpha\to 0$, see Eqs. (\ref{alpha
def}) and (\ref{alpha exp}), then the lack of inversion symmetry
does not have any appreciable effect on the upper critical field,
which is described by the standard Helfand-Werthamer theory.
However, as $\alpha$ increases, so do the deviations from the
centrosymmetric case. In the extreme paramagnetic limit,
corresponding to $\alpha\to\infty$, the critical field diverges at
$T\to 0$ in the clean case.

Impurities suppress both the orbital and paramagnetic pair
breaking, see Eqs. (\ref{universal eq}) and (\ref{univeral pair
breaking orbital}), resulting in an enhancement of the upper
critical field. In the strongly paramagnetic limit, the effects of
disorder on the low-temperature critical field are nonmonotonic:
At weak disorder $H_{c2}$ decreases, but in the ``dirty'' limit
the overall enhancement of the critical field takes over.

The spatial structure of the superconducting order parameter is,
in general, different from the centrosymmetric case. As soon as
the difference between the densities of states in the SO-split
bands, $N_+$ and $N_-$, is taken into account, the system exhibits
a helical instability, in which the order parameter has a phase
gradient along the applied magnetic field (in a cubic crystal).
The helical superconducting state turns out to be robust with
respect to both the orbital pair breaking and disorder. In
particular, the wave vector of the superconducting instability in
in the Ginzburg-Landau regime and in the ``dirty'' limit is given
by $q_z=-2\delta\alpha(H/H_0)\xi_0^{-1}$. In contrast, the LOFF
state, which occupies the low-temperature part of the phase
diagram at $N_+=N_-$, is quickly suppressed by both the orbital
interaction and disorder, similarly to its counterpart in the
centrosymmetric case.\cite{GG66,Aslam68}

Let us discuss the application of our results to the family of
cubic noncentrosymmetric compounds Li$_2$(Pd$_{1-x}$,Pt$_x$)$_3$B,
where $x$ ranges from 0 to 1. The critical temperature varies from
7-8 K for $x=0$ to 2.2-2.8 K for $x=1$. The electronic band
structure also exhibits considerable variation: The SO band
splitting is strongly anisotropic, and can be as large as 30 meV
in Li$_2$Pd$_3$B, and 200 meV in Li$_2$Pt$_3$B (Ref.
\onlinecite{LP05}). Experimental data\cite{Yuan-Nishi,Hafl07} seem
to agree that Li$_2$Pd$_3$B is a conventional BCS-like
superconductor with no gap nodes. In contrast, the gap structure
in Li$_2$Pt$_3$B is still a subject of controversy. While earlier
experiments, see Refs. \onlinecite{Yuan-Nishi}, suggested the
presence of lines of nodes in the gap, the recent $\mu$SR and
specific heat data\cite{Hafl07} have found no evidence of those.
Moreover, according to Ref. \onlinecite{Hafl07}, the whole
Li$_2$(Pd$_{1-x}$,Pt$_x$)$_3$B family of compounds are single-gap
isotropic superconductors. If this is the case then our model
based on a BCS-contact pairing Hamiltonian should be applicable.
In these compounds, the paramagnetic effects seem to be rather
weak: the parameter $\alpha$ varies from 0.10 for $x=0$ to $0.15$
for $x=1$ (using the data from Ref. \onlinecite{Hafl07}). This
explains why the experimental $H_{c2}$ curves in these materials
show a good agreement with the Helfand-Werthamer theory. Still,
the absence of inversion symmetry should manifest itself in a
long-wavelength helical modulation of the order parameter along
the applied field. Using the maximum values of the SO band
splitting from Ref. \onlinecite{LP05}, we have: $E_{SO}\simeq 30$
meV in Li$_2$Pd$_3$B, and $E_{SO}\simeq 200$ meV in Li$_2$Pt$_3$B.
Then one can obtain the following order-of-magnitude estimates for
the wave vector of the helical modulation: $q_z\xi_0\sim
10^{-3}H/H_{c2}(0)$ in Li$_2$Pd$_3$B, and $q_z\xi_0\sim
10^{-2}H/H_{c2}(0)$ in Li$_2$Pt$_3$B.

\section*{Acknowledgments}

The author is pleased to thank V. P. Mineev for stimulating
discussions. The financial support from the Natural Sciences and
Engineering Research Council of Canada is gratefully acknowledged.

\appendix

\section{Spectrum of $\hat{\mathcal{Y}}_{00}(\omega_n)$}

The problem of finding the upper critical field is reduced to the
calculation of the average normal-state Green's function in a
uniform magnetic field. The difference from the centrosymmetric
case is due to the presence of the SO coupling term in the
Hamiltonian (\ref{hat h}). Following Ref. \onlinecite{Gor59}, we
represent the Green's function before disorder averaging in a
factorized form:
\begin{equation}
\label{GF factorized}
    G_{\alpha\beta}(\br_1,\br_2;\omega_n)=
    \bar G_{\alpha\beta}(\br_1,\br_2;\omega_n)e^{i\varphi(\br_1,\br_2)},
\end{equation}
where
$\varphi(\br_1,\br_2)=(e/c)\int_{\br_1}^{\br_2}\bm{A}(\br)d\br$,
and the integration is performed along a straight line connecting
$\br_1$ and $\br_2$. From Eq. (\ref{G eq gen}) we obtain an
equation for the gauge-invariant factor:
\begin{equation}
\label{bar G eq gen}
    \bigl(i\omega_n-\hat{\bar{h}}_1\bigr)
    \hat{\bar{G}}(\br_1,\br_2;\omega_n)=\delta(\br_1-\br_2),
\end{equation}
where
\begin{equation}
\label{bar H gen}
    \hat{\bar{h}}_1=\frac{\bar\bK_1^2}{2m^*}+\gamma_0\bar\bK_1\bsigma
    +\frac{g}{2}\mu_B\bH\bsigma+U(\br_1)-\epsilon_F,
\end{equation}
and $\bar\bK_1=-i\bnab_1+(e/2c)\bH\times(\br_1-\br_2)$.

As discussed in Sec. \ref{sec: gap equation}, in the limit of weak
SO coupling it is sufficient to consider only the singlet-singlet
terms in the Cooper impurity ladder. After disorder averaging, the
kernel (\ref{X_0}) takes the following form:
\begin{equation}
\label{X_0 G bar}
    \mathcal{Y}_{00}(\br_1,\br_2;\omega_n)=
    \bar{\mathcal{Y}}_{00}(\br_1-\br_2,\omega_n)e^{2i\varphi(\br_1,\br_2)},
\end{equation}
where the translationally-invariant part is defined by its Fourier
transform as follows:
\begin{eqnarray}
\label{bar X_0 q}
    &&\bar{\mathcal{Y}}_{00}(\bq,\omega_n)=\frac{1}{2\pi N_F}
    \nonumber\\
    &&\ \times\int\frac{d^3\bk}{(2\pi)^3}\tr[\hat{\bar G}(\bk+\bq,\omega_n)\hat\sigma_2
    \hat{\bar G}^T(-\bk,-\omega_n)\hat\sigma_2].\qquad
\end{eqnarray}
The Green's functions here are disorder-averaged solutions of Eq.
(\ref{bar G eq gen}). The next step is to use the identity
$e^{2i\varphi(\br_1,\br_2)}\eta(\br_2)=e^{-i(\br_1-\br_2)\bD_1}\eta(\br_1)$,
where $\bD=-i\bnab+(2e/c)\bm{A}$, from which it follows that the
operator $\hat{\mathcal{Y}}_{00}(\omega_n)$ can be obtained from
$\bar{\mathcal{Y}}_{00}(\bq,\omega_n)$ by replacing ${\bq\to\bD}$.
The gauge-invariant part of the Green's function contains the
effects of the SO coupling, the Zeeman interaction, the orbital
Landau quantization, and disorder. We treat the magnetic field
effects on $\hat{\bar G}$ perturbatively, which is legitimate if
the Zeeman energy is small compared with the SO band splitting,
i.e. $(g/2)\mu_BH\ll E_{SO}$, and also if the temperature is not
very low, so that the Landau level quantization can be neglected.

In the clean case, keeping only the terms linear in $H$ in Eq.
(\ref{bar G eq gen}), we have
\begin{equation}
\label{bar H K}
    \bigl(i\omega_n-\hat
    h_0+\bH\hat{\bm{m}}\bigr)\hat{\bar{G}}(\br,\omega_n)=\delta(\br),
\end{equation}
where $\hat h_0=\epsilon_0(-i\bnab)+\bgam(-i\bnab)\bsigma$ is the
zero-field Hamiltonian, and
$\hat{\bm{m}}=-\mu_B[(g/2)\bsigma+(\bR\times\hat{\bm{\pi}})]$,
with $\hat{\bm{\pi}}=(m/m^*)(-i\bnab)+m\gamma_0\bsigma$, has the
meaning of the magnetic moment operator for band electrons. The
eigenvalues of $\hat h_0$ are given by $\xi_\lambda(\bk)$, see
Eqs. (\ref{xis}). It is straightforward to show that the linear in
$H$ effects on $\hat{\bar G}$ are due to the Zeeman interaction
(the first term in $\hat{\bm{m}}$). Then,
$\hat{\bar{G}}(\bk,\omega_n)$ can be represented in the form
(\ref{G clean zero H}), with the band Green's functions now given
by
\begin{equation}
\label{G lambda clean}
    G_\lambda(\bk,\omega_n)=
    \frac{1}{i\omega_n-\xi_\lambda(\bk)-\lambda(g/2)\mu_B\hat\bk\bH}.
\end{equation}
Thus, the electron bands are deformed by the magnetic field:
$\xi_\lambda(\bk)\to\xi_\lambda(\bk)+\lambda(g/2)\mu_B\hat\bk\bH$.

In the presence of impurities, again keeping the magnetic field
only in the Zeeman term in Eq. (\ref{bar H gen}), we obtain for
the average Green's function:
\begin{equation}
\label{G average}
    \hat{\bar{G}}(\bk,\omega_n)=[i\omega_n-\epsilon_0(\bk)-\tilde{\bgam}(\bk)\bsigma
    -\hat\Sigma_{imp}(\omega_n)]^{-1},
\end{equation}
where $\tilde{\bgam}(\bk)=\gamma_0\bk+(g/2)\mu_B\bH$, and
\begin{equation}
\label{Sigma_imp}
    \hat\Sigma_{imp}(\omega_n)=n_{imp}U_0^2\int\frac{d^3\bk}{(2\pi)^3}
    \hat{\bar{G}}(\bk,\omega_n)
\end{equation}
is the impurity self-energy in the self-consistent Born
approximation. We seek the self-energy matrix in a spin-diagonal
form: $\hat\Sigma_{imp}(\omega_n)=-i\beta(\omega_n)\hat\sigma_0$.
Then,
$\hat{\bar{G}}(\bk,\omega_n)=\sum_{\lambda}\hat\Pi_\lambda(\bk)G_\lambda(\bk,\tilde\omega_n)$,
where $\tilde\omega_n=\omega_n+\beta(\omega_n)$, and the band
projection operators and the band Green's function are given by
expressions (\ref{Pis}) and (\ref{G lambda clean}) respectively.
Substituting this into Eq. (\ref{Sigma_imp}) and neglecting the
energy dependence of the band densities of states on the scale of
the Zeeman energy, we find:
$\tilde\omega_n=\omega_n+\Gamma\sign\omega_n$. Therefore,
\begin{eqnarray}
\label{bar G average}
    &&\hat{\bar{G}}(\bk,\omega_n)=\sum_{\lambda}\hat\Pi_\lambda(\bk)\nonumber\\
    &&\quad\times\frac{1}{i\omega_n-\xi_\lambda(\bk)-\lambda(g/2)\mu_B\hat\bk\bH
    +i\Gamma\sign\omega_n}.\qquad
\end{eqnarray}

Inserting the last expression in Eq. (\ref{bar X_0 q}), we obtain:
\begin{eqnarray}
\label{bar X_0 Omega}
    &&\bar{\mathcal{Y}}_{00}(\bq,\omega_n)\nonumber\\
    &&\quad =\frac{1}{2}\sum_\lambda\rho_\lambda\left\langle
    \frac{1}{|\omega_n|+\Gamma+i\Omega_\lambda\sign\omega_n}
    \right\rangle_{\hat{\bk}},\quad
\end{eqnarray}
where $\rho_\lambda=N_\lambda/N_F$, and
$\Omega_{\lambda}(\bk,\bq)=v_F\hat\bk(\bq+\lambda
g\mu_B\bH/v_F)/2$. Next we use in Eq. (\ref{bar X_0 Omega}) the
identity $a^{-1}=\int_0^\infty du\,e^{-au}$ and make the
substitution $\bq\to\bD$ to represent
$\hat{\mathcal{Y}}_{00}(\omega_n)$ as a differential operator of
infinite order:
\begin{equation}
\label{hat X_0 final}
    \hat{\mathcal{Y}}_{00}(\omega_n)=\frac{1}{2}\int_0^\infty du\;e^{-u(|\omega_n|+\Gamma)}
    \sum_\lambda\rho_\lambda\hat{\cal O}_\lambda.
\end{equation}
Here
\begin{equation}
\label{O def}
    \hat{\cal O}_\lambda=\left\langle\exp
    \left[-\frac{iuv_F}{2}\hat{\bk}\left(\bD+\lambda\frac{g\mu_B\bH}{v_F}
    \right)\right]\right\rangle_{\hat{\bk}}
\end{equation}
do not depend on $\omega_n$, because the zero-field Fermi surfaces
are invariant under $\bk\to-\bk$.

To find the eigenfunctions and eigenvalues of $\hat{\cal
O}_\lambda$, we follow the procedure outlined in Ref.
\onlinecite{HW66}. We choose the $z$-axis along the external
field, so that $\bH=H\hat z$, and introduce the operators
\begin{equation}
\label{a operators}
    a_\pm=\ell_H\frac{D_x\pm iD_y}{2},\quad
    a_3=\ell_HD_z,
\end{equation}
where $\ell_H=\sqrt{c/eH}$ is the magnetic length. It is easy to
check that $a_+=a_-^\dagger$ and $[a_-,a_+]=1$, therefore $a_\pm$
have the meaning of the raising and lowering operators, while
$a_3=a_3^\dagger$ commutes with both of them: $[a_3,a_\pm]=0$. We
use the basis of states $|N,p\rangle$ (Landau levels), such that
\begin{eqnarray*}
    &&a_+|N,p\rangle=\sqrt{N+1}|N+1,p\rangle\\
    &&a_-|N,p\rangle=\sqrt{N}|N-1,p\rangle\\
    &&a_3|N,p\rangle=p|N,p\rangle,
\end{eqnarray*}
where $N=0,1,...$, and $p$ is a real number. Then,
\begin{eqnarray}
    \hat{\cal O}_\lambda|N,p\rangle=\frac{1}{2}\int_0^\pi
    d\theta\sin\theta\,e^{-iv(p+\lambda
    p_0)\cos\theta}\nonumber\\
    \times\int_0^{2\pi}\frac{d\phi}{2\pi}e^{-iv\sin\theta
    (e^{-i\phi}a_++e^{i\phi}a_-)}|N,p\rangle\nonumber\\
    =\frac{1}{2}\int_{-1}^1 ds\,e^{-iv(p+\lambda p_0)s}e^{-(v^2/2)(1-s^2)}
    \nonumber\\
    \times L_N[v^2(1-s^2)]|N,p\rangle,
\end{eqnarray}
where $v=(v_F/2\ell_H)u$, $p_0=g\mu_BH\ell_H/v_F$, and $L_N(x)$ is
the Laguerre polynomial of degree $N$. We see that the Landau
levels $|N,p\rangle$ are eigenfunctions of $\hat{\cal O}_\lambda$
and, therefore, of $\hat{\mathcal{Y}}_{00}(\omega_n)$. Introducing
$\delta=(\rho_+-\rho_-)/2$ and summing over $\lambda$ in Eq.
(\ref{hat X_0 final}), we arrive at the expression (\ref{y N p}).


\begin{thebibliography}{99}


\bibitem{Bauer04}
E. Bauer, G. Hilscher, H. Michor, Ch. Paul, E. W. Scheidt, A.
Gribanov, Yu. Seropegin, H. No\"el, M. Sigrist, and P. Rogl, Phys.
Rev. Lett. \textbf{92}, 027003 (2004).

\bibitem{LiPt-PdB}
K. Togano, P. Badica, Y. Nakamori, S. Orimo, H. Takeya, and K.
Hirata, Phys. Rev. Lett. \textbf{93}, 247004 (2004); P. Badica, T.
Kondo, and K. Togano, J. Phys. Soc. Jpn. \textbf{74}, 1014 (2005).

\bibitem{Edel89}
V. M. Edelstein, Zh. Eksp. Teor. Fiz. \textbf{95}, 2151 (1989)
[Sov. Phys. JETP \textbf{68}, 1244 (1989)].

\bibitem{GR01}
L. P. Gor'kov and E. I. Rashba, Phys. Rev. Lett. \textbf{87},
037004 (2001).

\bibitem{Yip02}
S. K. Yip, Phys. Rev. B \textbf{65}, 144508 (2002).

\bibitem{FAKS04}
P. A. Frigeri, D. F. Agterberg, A. Koga, and M. Sigrist, Phys.
Rev. Lett. \textbf{92}, 097001 (2004) [Erratum \textbf{93},
099903(E) (2004)].

\bibitem{FAS04}
P. A. Frigeri, D. F. Agterberg, and M. Sigrist, New J. Phys.
\textbf{6}, 115 (2004).

\bibitem{Sam05}
K. V. Samokhin, Phys. Rev. Lett. \textbf{94}, 027004 (2005).

\bibitem{Sam07}
K. V. Samokhin, Phys. Rev. B \textbf{76}, 094516 (2007).

\bibitem{Lev85}
L. S. Levitov, Yu. V. Nazarov, and G. M. Eliashberg, Pis'ma Zh.
Eksp. Teor. Fiz. \textbf{41}, 365 (1985) [JETP Letters
\textbf{41}, 445 (1985)].

\bibitem{Edel95}
V. M. Edelstein, Phys.  Rev.  Lett. \textbf{75}, 2004 (1995).

\bibitem{Fuji05}
S. Fujimoto, Phys. Rev. B \textbf{72}, 024515 (2005).

\bibitem{Agter03}
D. F. Agterberg, Physica C \textbf{387}, 13 (2003).

\bibitem{Sam04}
K. V. Samokhin, Phys. Rev. B \textbf{70}, 104521 (2004).

\bibitem{KAS05}
R. P. Kaur, D. F. Agterberg, and M. Sigrist, Phys. Rev. Lett.
\textbf{94}, 137002 (2005).

\bibitem{HW66}
E. Helfand and N. R. Werthamer, Phys. Rev. \textbf{147}, 288
(1966).

\bibitem{WHH66}
N. R. Werthamer, E. Helfand, and P. C. Hohenberg, Phys. Rev.
\textbf{147}, 295 (1966).

\bibitem{Scho91}
C. T. Rieck, K. Scharnberg, and N. Schopohl, J. Low Temp. Phys. 
\textbf{84}, 381 (1991).

\bibitem{BG02}
V. Barzykin and L. P. Gor'kov, Phys. Rev. Lett. \textbf{89},
227002 (2002).

\bibitem{DF07}
O. Dimitrova and M. V. Feigel'man, Phys. Rev. B \textbf{76},
014522 (2007).

\bibitem{MS07}
V. P. Mineev and K. V. Samokhin, Phys. Rev. B \textbf{75}, 184529
(2007).

\bibitem{Rashba60}
E. I. Rashba, Fiz.  Tverd.  Tela (Leningrad) \textbf{2}, 1224
(1960) [Sov. Phys. Solid State \textbf{2}, 1109 (1960)].

\bibitem{SM08}
K. V. Samokhin and V. P. Mineev, Phys. Rev. B \textbf{77}, 104520
(2008).

\bibitem{LL9}
E. M. Lifshitz and L. P. Pitaevskii, \emph{Statistical Physics,
Part 2} (Butterworth-Heinemann, Oxford, 1995).

\bibitem{AGD}
A. A. Abrikosov, L. P. Gorkov, and I. E. Dzyaloshinski,
\emph{Methods of Quantum Field Theory in Statistical Physics}
(Dover, New York, 1975).

\bibitem{Sam08}
K. V. Samokhin, to be published.

\bibitem{Maki64}
K. Maki, Physics \textbf{1}, 127 (1964).

\bibitem{Gor60}
L. P. Gor'kov, Zh. Eksp. Teor. Fiz. \textbf{37}, 1407 (1959) [Sov.
Phys. -- JETP \textbf{10}, 998 (1960)].

\bibitem{LOFF}
A. I. Larkin and Yu. N. Ovchinnikov, Zh. Eksp. Teor. Fiz.
\textbf{47}, 1136 (1964) [Sov. Phys. JETP \textbf{20}, 762
(1965)]; P. Fulde and R. A. Ferrell, Phys. Rev. \textbf{135}, A550
(1964).

\bibitem{AK07}
D. F. Agterberg and R. P. Kaur, Phys. Rev. B \textbf{75}, 064511
(2007).

\bibitem{divergence}
One should keep in mind that Eq. (\ref{w Q paramag}) and
subsequent results are valid only if the Zeeman energy is small
compared with the SO band splitting, i.e. at $\tilde h\ll
E_{SO}/T_{c0}$.

\bibitem{Tink-book}
M. Tinkham, \emph{Introduction to Superconductivity}, Ch. 10.2
(McGraw-Hill, New York, 1996).

\bibitem{CC62}
A. M. Clogston, Phys. Rev. Lett. \textbf{9}, 266 (1962); B. S.
Chandrasekhar, Appl. Phys. Lett. \textbf{1}, 7 (1962).

\bibitem{GG66}
L. W. Gruenberg and L. Gunther, Phys. Rev. Lett. \textbf{16}, 996
(1966).

\bibitem{Aslam68}
L. G. Aslamazov, Zh. Eksp. Teor. Fiz. \textbf{55}, 1477 (1968)
[Sov. Phys. -- JETP \textbf{28}, 773 (1969)].

\bibitem{LP05}
K.-W. Lee and W. E. Pickett, Phys. Rev. B \textbf{72}, 174505
(2005).

\bibitem{Yuan-Nishi}
H. Q. Yuan, D. F. Agterberg, N. Hayashi, P. Badica, D.
Vandervelde, K. Togano, M. Sigrist, and M. B. Salamon, Phys. Rev.
Lett. \textbf{97}, 017006 (2006); M. Nishiyama, Y. Inada, and
G.-Q. Zheng, Phys. Rev. Lett. \textbf{98}, 047002 (2007).

\bibitem{Hafl07}
P. S. H\"afliger, R. Khasanov, R. Lortz, A. Petrovi\'c, K. Togano,
C. Baines, B. Graneli, and H. Keller, preprint arXiv:0709.3777
(unpublished).

\bibitem{Gor59}
L. P. Gor'kov, Zh. Eksp. Teor. Fiz. \textbf{36}, 1918 (1959) [Sov.
Phys. -- JETP \textbf{9}, 1364 (1959)].

\end{thebibliography}
\end{document}